\documentclass{article}

\usepackage{arxiv}
\usepackage{float}
\usepackage[utf8]{inputenc} 
\usepackage[T1]{fontenc}    
\usepackage{hyperref}       
\usepackage{url}            
\usepackage{booktabs}       
\usepackage{amsfonts}       
\usepackage{nicefrac}       
\usepackage{microtype}    
\usepackage{lipsum}		   
\usepackage{graphicx}
\usepackage{natbib}
\usepackage{doi}
\usepackage{threeparttable} 
\usepackage{siunitx}
\usepackage{multirow}
\usepackage{enumitem}
\usepackage{array}

\newcommand{\authorc}{\citep{khosravi_development_2020}} 
\newcommand{\authord}{\citep{lahza_effects_2022}} 
\newcommand{\authore}{\citep{gyamfi_effects_2022}} 
\newcommand{\authorf}{\citep{darvishi2022incorporating}} 
\newcommand{\authorg}{\citep{darvishi_impact_2024}} 
\title{Emotionally Enriched Feedback via Generative AI}

\renewcommand{\shorttitle}{Emotionally Enriched AI Feedback}

\hypersetup{
pdftitle={Emotionally Enriched Feedback via Generative AI},
pdfsubject={Educational Technology, Artificial Intelligence},
pdfauthor={Omar Alsaiari, Nilufar Baghaei, Hatim Lahza, Jason M. Lodge, Marie Boden, Hassan Khosravi},
pdfkeywords={Generative AI Feedback, Emotional Responses, Student Engagement, Control-Value Theory},
}

\author{
  Omar Alsaiari \\
  School of Electrical Engineering and Computer Science \\
  The University of Queensland \\
  St Lucia, QLD, 4072, Australia \\
  \texttt{o.alsaiari@uq.net.au} \\
  \And
  Nilufar Baghaei \\
  School of Electrical Engineering and Computer Science \\
  The University of Queensland \\
  St Lucia, QLD, 4072, Australia \\
  \texttt{n.baghaei@uq.edu.au} \\
  \And
  Hatim Lahza \\
  School of Electrical Engineering and Computer Science \\
  The University of Queensland \\
  St Lucia, QLD, 4072, Australia \\
  \texttt{h.lahza@uq.net.au} \\
  \And
  Jason M. Lodge \\
  School of Education \\
  The University of Queensland \\
  St Lucia, QLD, 4072, Australia \\
  \texttt{j.lodge@uq.edu.au} \\
  \And
  Marie Boden \\
  School of Electrical Engineering and Computer Science \\
  The University of Queensland \\
  St Lucia, QLD, 4072, Australia \\
  \texttt{marieb@eecs.uq.edu.au} \\
  \And
  Hassan Khosravi \\
  Institute for Teaching and Learning Innovation\\
  The University of Queensland \\
  St Lucia, QLD, 4072, Australia \\
  \texttt{h.khosravi@uq.edu.au} \\
}

\begin{document}
\maketitle

\begin{abstract}
This study investigates the impact of emotionally enriched AI feedback on student engagement and emotional responses in higher education. Leveraging the Control-Value Theory of Achievement Emotions, we conducted a randomized controlled experiment involving 425 participants where the experimental group received AI feedback enhanced with motivational elements, while the control group received neutral feedback. Our findings reveal that emotionally enriched feedback is perceived as more beneficial and helps reduce negative emotions, particularly anger, towards receiving feedback. However, it had no significant impact on the level of engagement with feedback or the quality of student work. These results suggest that incorporating emotional elements into AI-driven feedback can positively influence student perceptions and emotional well-being, without compromising work quality. Our study contributes to the growing body of research on AI in education by highlighting the importance of emotional considerations in educational technology design.
\end{abstract}


\section{Introduction \label{sec:intro}}

The importance of feedback in learning has been widely acknowledged, with much research and scholarship focusing on identifying criteria for effective feedback, including being detailed, actionable, and constructive \citep{doi:10.1080/13562517.2022.2029394}.  Feedback processes also go far beyond the mere provision of information. Jensen and colleagues \citep{doi:10.1080/02602938.2022.2059446} discuss the notion of 'feedback encounters' as a way of encapsulating the complicated nature of feedback in higher education settings. High-quality feedback is, therefore, critical but also complicated.

The need for effective feedback becomes even more critical in the context of online learning with large student enrolments. In such settings, engaging students in timely and personalized feedback is a significant challenge for educators across modes of study. The sheer volume of students can make it impractical to ensure that high-quality feedback is consistently available.  Feedback encounters in this context can become transactional, going against contemporary notions of effective feedback that are relational and involve multiple processes such as student sensemaking and planning \citep{doi:10.1080/02602938.2012.691462}.  

 A further element of feedback encounters that can be lost when shifting away from physical learning environments is difficulty in communicating and reading emotion. Many of the usual cues available in a physical setting, such as facial expressions are absent. This lack of emotion is critical in feedback because the triggering of emotional responses during feedback encounters will contribute to determining whether learners successfully adjust their strategies \citep{doi:10.1080/00461520.2022.2134135}. As such, there are difficult challenges evident in online and other digital learning environments for engaging students in effective feedback encounters. 
 
 The control-value theory (CVT) of achievement emotions \citep{pekrun2007control} provides a robust framework for understanding the role of emotion in feedback processes. Positive emotions, such as enjoyment and hope, have been shown to enhance motivation and academic success, while negative emotions, like anger and anxiety, can hinder these outcomes. By incorporating emotional elements such as encouragement, praise, and empathetic language into feedback encounters facilitated by AI in a digital learning environment, we hypothesize that students will experience reduced negative emotions and increased engagement. The study is an attempt to deploy emerging AI technologies to help overcome some of the difficulties in engaging students in effective feedback in digital environments.

This study aims to investigate the potential of generative AI to enhance engagement with feedback processes through emotional enrichment. Our research is grounded in the CVT of achievement emotions, which posits that students' emotions in academic settings are influenced by their perceived control over learning activities and the value they attach to these activities. We conducted a field-based randomized controlled experiment in a first-year engineering course with 425 students. In this course, students engaged in co-creation by developing study resources (e.g., multiple-choice questions) and providing peer feedback on each other's creations. Large language models (LLMs) were used to deliver instant feedback on both the students' created content and their peer feedback. The control group received feedback in a neutral tone. In contrast, the experimental group received feedback generated from the same prompt, but with the LLM instructed to incorporate positive elements such as praise, encouragement, and visual aids like emojis, all conveyed in an overall positive and uplifting tone. The impact of this intervention will be evaluated through a mixed-methods approach, assessing various measures including students' level of engagement with feedback, their perception of its effectiveness, their emotional response to receiving feedback, and the impact on the quality of their work.
Our research and study is guided by the following research questions.

\begin{itemize}
    \item \noindent \textbf{RQ1. Engagement with Feedback:} What is the impact of AI emotionally enriched feedback on student engagement with feedback processes? A critical challenge with feedback is that students often do not engage with it .Investigating the extent to which students engage with feedback is crucial because it highlights the effectiveness of the feedback in fostering active learning and participation, which are key to academic success \citep{winstone2017supporting}.  Here we explore whether the experimental group would demonstrate more engagement with the provided feedback compared to the control group.

    \item \noindent \textbf{RQ2. Usefulness perception:} How do students perceive the usefulness of AI emotionally enriched feedback? Understanding students' perceptions of the feedback is essential as it influences their motivation to act on it. If students find the feedback valuable, they are more likely to incorporate it into their learning process \citep{gamlem2013student}. Here we explore whether the experimental group will perceive the AI-based feedback to be more beneficial compared to the control group.
    \item \noindent \textbf{RQ3. Emotional response to feedback:} What is the effect of AI emotionally enriched feedback on students' emotional responses to feedback? The emotional response to feedback can significantly impact students' attitudes towards learning and their willingness to engage with educational content. Exploring the emotional responses to feedback is essential as it sheds light on how feedback influences students' emotional well-being and learning engagement   \citep{d2016emotions}. Here we explore how emotionally enriched AI feedback will impact students' positive emotions such as (a) enjoyment, (b) hope, and (c) pride 
 as well as negative emotions such as (a) anger, (b) anxiety, (c) shame, (d) hopelessness, and (e) boredom compared to students in the control group.
 \item \noindent \textbf{RQ4: Impact on outcome} How does AI emotionally enriched feedback affect the quality of student work? Ultimately, the aim of feedback is to contribute to better learning outcomes. Understanding the impact of feedback on the quality of student work is essential for assessing its overall effectiveness. Here we explore whether students receiving emotionally enriched feedback will produce higher quality resources compared to those receiving neutral feedback \citep{Faulconer2021The}. 
\end{itemize}

The structure of this paper is as follows: Section \ref{sec:lit} provides an overview of the related literature. Section \ref{sec:method} outlines the methodology used for the study. Section \ref{sec:Results} presents the results, and Section \ref{sec:Dis} discusses the findings in relation to existing literature, offering likely explanations for the outcomes.
The insights gained from the study aim to contribute to the broader discourse on educational technologies by emphasizing the significance and aspiring to create a more holistic learning experience that fosters academic success and emotional well-being.

\section{Literature review}\label{sec:lit}

\subsection{Artificial Intelligence in Education (AIEd)}
Artificial Intelligence (AI) in education represents a transformative force, encompassing a wide array of technologies such as machine learning, natural language processing, data mining, and neural networks. The integration of AI into educational settings offers numerous benefits, including personalized learning experiences, adaptive learning technologies, and the automation of administrative tasks. These advancements have the potential to enhance the educational process by catering to individual student needs, optimizing resource allocation, and providing real-time feedback \citep{holmes2023artificial, chen2020artificial}.

One of the key advantages of AI in education is its ability to provide personalized learning experiences. AI systems can analyze students' learning patterns, strengths, and weaknesses, thereby customizing educational content to meet their specific needs. This approach not only improves engagement and motivation but also enhances learning outcomes by allowing students to progress at their own pace \citep{zawacki2019}.

Adaptive learning technologies, another critical component of AI in education, use algorithms to adjust the difficulty of educational tasks based on student performance. This dynamic adjustment ensures that students are continually challenged at an appropriate level, which promotes better understanding and retention of the material \citep{chen2020artificial}. Additionally, AI-driven administrative automation can streamline various educational processes, such as grading and scheduling, thus allowing educators to focus more on teaching and student interaction \citep{roll2016evolution}.

Generative AI, a subset of AI, is particularly noteworthy for its potential in education. It is poised to revolutionize educational feedback mechanisms by providing personalized and instantaneous responses to students' work.\citep{phung2023generative} demonstrate that GPT-4 can significantly outperform earlier models and nearly match human tutors in offering detailed feedback encounters in introductory programming problems, suggesting a promising future for AI-assisted learning environments. \citep{sharples2023towards} expands on this by exploring how generative AI can foster continuous dialogue between learners and AI, creating a dynamic feedback loop that enhances understanding and retention through interactive and adaptive feedback. This finding is critical as it points to the potential for AI to engage in feedback with students in a less transactional, more dialogic and relational manner. In a study by \citep{smolansky2023educator}, the impact of generative AI on educational assessments is examined, revealing that while educators appreciate the AI's ability to provide immediate, formative feedback that encourages critical thinking, students exhibit mixed reactions, reflecting concerns about creativity and authenticity. 
Additionally,\citep{darvishi2024} indicates that AI-generated feedback is often perceived as more detailed and helpful compared to traditional feedback, which can significantly aid in student learning and improvement. However, it is crucial to manage the use of AI to avoid over-reliance and ensure that students continue to develop critical thinking and self-regulation skills.

In conclusion, AI, including generative AI, holds immense potential to augment education by offering personalized, adaptive, and efficient learning experiences. As these technologies continue to evolve, it is essential to conduct empirical research to fully understand their impact and to develop best practices for their integration into educational frameworks.\citep{shaik2022review}.
\subsection{Enhancing Feedback through AI}
Feedback is portrayed as a collaborative process in which both educators and learners actively engage, exchanging insights, reflections, and guidance. This dynamic interaction, far from being a one-way communication from teachers to students, aims to promote self-regulation, encourage active learning, and develop students' evaluative skills within the educational setting \citep{doi:10.1080/02602938.2012.691462}. Feedback plays a pivotal role in both traditional and online learning contexts, serving as a key connector among learners, instructors, and peers \citep{kluger1996,sadler1989,shute2008}. In online environments, feedback extends beyond instructor comments, encompassing various forms of automatically generated information, including AI-generated feedback \citep{darvishi2024}.

Recent research indicates that AI can effectively engage students in feedback encounters, enhancing their learning experiences and outcomes. Studies have shown that AI-generated feedback, such as GPT-3.5, is often perceived as more useful than human feedback and requires minimal modification by instructors \citep{wan2023e}. AI-based approaches, including explainable AI and machine learning, have been proposed to support and improve students' self-regulation and learning achievements \citep{afzaal2023i}. These approaches have been applied in various contexts, including language learning, where tools like Grammarly have significantly improved writing performance \citep{chang2021e}.

However, alongside these positive impacts, research also points to the potential downsides of AI assistance in learning environments. A study on the impact of AI on student agency in online learning contexts revealed that while AI tools like real-time feedback systems can scaffold learning, they might also lead to an over-reliance by students. This dependency could adversely affect their ability to self-regulate and maintain agency in their learning processes \citep{darvishi2024}. These studies collectively suggest that AI-generated feedback can be a valuable tool in various educational contexts, offering advantages like immediate response, personalization, and scalability.

\subsection{The Control-Value Theory (CVT) of Achievement Emotions}
CVT of achievement emotions, developed by Pekrun, emphasizes the role of control and value appraisals in shaping students' emotional experiences in academic contexts. According to this theory, students' emotions are influenced by their perceived level of control over learning activities and the value they attach to these activities \citep{pekrun2006control}. The influence of emotions on academic achievement is intricate, shaped by both the type of emotion and the context in which it occurs. Positive activating emotions like enjoyment, which are linked to learning, tend to boost interest and strengthen both intrinsic and extrinsic motivation. Conversely, negative deactivating emotions such as hopelessness and boredom typically undermine motivation and are detrimental to performance. The impact of positive deactivating emotions, like relief, and negative activating emotions, including anger, anxiety, and shame, is more complex and varies significantly. For instance, while anxiety about failing an exam might reduce a student's intrinsic interest in the material, it can simultaneously increase their motivation to exert effort to avoid failure \citep{pekrun2002academic, pekrun2004beyond}.

Research supports these dynamics, showing that emotions significantly influence academic outcomes, including through feedback processes \cite{doi:10.1080/00461520.2022.2134135}. For instance, enjoyment of learning not only enhances motivation but also correlates positively with academic success. In contrast, emotions like hopelessness and boredom are associated with negative academic outcomes. The effects of anxiety and shame are mixed, showing both positive and negative influences on motivation and achievement, highlighting their complex role in educational contexts \citep{pekrun2007control, pekrun2002academic}. This intricate relationship suggests that emotions play a critical and nuanced role in shaping students' academic trajectories.

\subsection{The Role of Feedback in Shaping Emotions and Academic Outcomes}
Feedback on success and failure significantly influences students' emotions that are directly linked to their academic achievements. Such feedback not only impacts their current emotional state but also shapes their expectations and perceived values of their future academic performances, which in turn determine their future emotional responses. Feedback, including messages from teachers about the reasons behind students' performances, helps shape students’ perceptions of their ability to control and value their performance outcomes. This is crucial for forming their future emotional reactions and assessments.

Studies have shown that how students perceive feedback significantly influences their emotional responses, impacting their academic behavior and results \citep{yang2019exploring}. When students doubt the validity of the feedback provided by teachers, they tend to feel upset and let down \citep{sargeant2008understanding}. Conversely, students often experience feelings of pleasure when they view feedback as constructive and useful for their progress, even if the feedback is critical \citep{fong2018inside}. Feelings of shame can arise from constructive feedback if students interpret it as an indication of their failure, or if they receive praise that they feel is not merited \citep{fong2018inside}.
These insights reveal that the type of emotional response from students depends less on the feedback itself and more on how they interpret it. Students who are more accepting and open to feedback typically view it more positively, which leads to favorable emotional reactions. Such positive emotions are essential as they lead to better learning outcomes \citep{frondozo2023feedback}. Fredrickson notes that positive emotions help broaden an individual's perspective, enabling them to see situations more clearly, enhance their problem-solving skills, and increase their capacity to manage adversity \citep{fredrickson2001role}. Consequently, students who harbor positive academic emotions are more likely to actively seek feedback and utilize it to seek solutions and enhance their performance \citep{yang2023unpacking}.

\section{Research methods}\label{sec:method}

\subsection{Research tool: The RiPPLE System \label{sec:system}}

We utilized RiPPLE, an adaptive educational system that employs a constructivist epistemology-driven learnersourcing \cite{khosravi2023learnersourcing}mechanism to engage students in socio-cognitive and socio-cultural-based learning through three main activities: content authoring, peer review, and drill and practice. Constructivism holds several assumptions; central and common among constructivists is the belief that learners have an active role in constructing their own learning \citep{peterson_constructivism_2010}. In the initial stage of the learnersourcing mechanism in RiPPLE, students create various learning resources including, but not limited to, worked examples, general notes, and multiple-choice questions (MCQs). This activity takes students through several steps that engage them cognitively with course materials. For example, when creating MCQs, students draft question content, tag relevant topics, generate potential answers and plausible distractors, and formulate explanations and rationales regarding correct and incorrect answers.

Moving forward, the learnersourcing mechanism considers the socio-cultural aspect, such as Vygotsky’s theory, which posits that social interaction and environment are essential for cognitive development \citep{vygotsky_mind_1978}. Social interaction can occur in various forms, such as student-to-peer, student-to-teacher, or student-to-machine (e.g., computer-based scaffolding) (see \citep{nardo_exploring_2021}). In RiPPLE, through a machine learning-enhanced peer review process, students, their peers, and instructors collaboratively refine the created learning resources to meet the quality standards provided on the platform. Furthermore, the learnersourcing mechanism also considers the socio-cognitive aspect \citep{bandura_social_1999}, which, similar to the socio-cultural perspective, stresses the role of environmental factors as part of a broader reciprocal interaction system that includes the environment, behaviours, cognition, and personal factors \citep{schunk_social_1989}. From this viewpoint, learning experiences can take two forms: enactive when performing tasks or vicarious when observing models \citep{schunk_social_1989}. In RiPPLE, students engage in vicarious learning when they review their peers' work and when they participate in drill and practice sessions (also seen as retrieval-based learning \citep{byrne_227_2017}), in which they answer questions or view notes and worked examples created by their peers.

\begin{figure}[t]
    \centering
    \includegraphics[width=.7\textwidth]{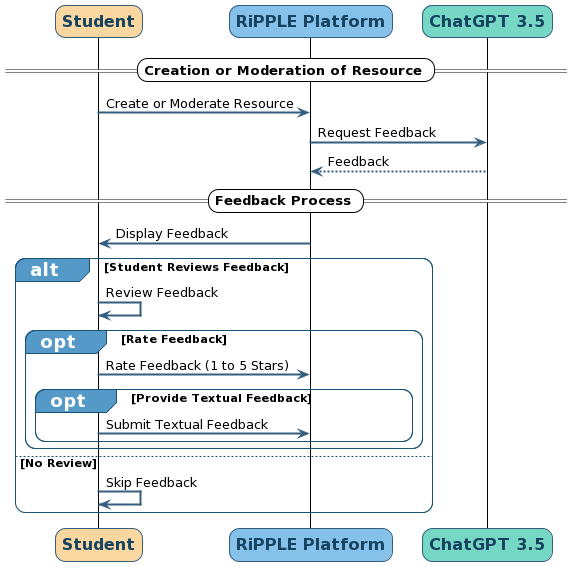} 
    \caption{UML Diagram: RiPPLE's AI Feedback Process and Student Rating Mechanism}
    \label{fig:UML}
    \begin{threeparttable}

    \end{threeparttable}
\end{figure}

As the learning content authored by students varies in quality, a significant line of research in learnersourcing aims to control the quality of students' contributions. Hence, while RiPPLE does not follow a strict design-based research (DBR) approach, it has several endeavours grounded in various literatures, including crowdsourcing, adaptive learning environments, learning sciences, and educational technology, to improve the moderation mechanism from the perspectives of content creators and content reviewers. An example of a study that aimed to control the quality of content authored by students is \authord’s work, where different self-regulated learning (SRL) scaffolds were provided to content creators to enhance their content quality and engagement (authord). On the other hand, \authore conducted a controlled experiment to investigate the effectiveness of incorporating proficiency-based or data-driven rubrics and criteria to improve the quality of written feedback provided by students via the peer review activity (\authore). In addition, \authorf has investigated the impact of involving instructors through an AI-based spot-checking feature to review only a subset of resources created by students \authorf. Moreover, the platform has recently integrated AI feedback \authorg. This feedback is currently used before either an authored resource or peer-written feedback is submitted (Figure \ref{fig:UML}). 

It is important to note that not all learning resources are accepted. Some resources that do not pass an AI-based consensus algorithm are denied. Therefore, learning resources that pass the moderation process are added to course repositories and become available for others in the course to use, attempt, and provide feedback on. Additionally, the platform aims to motivate students to engage more and value their role in the platform by allowing them to rate and comment on any of the learning resources and by assigning them ability ratings based on their performance on the platform, which are presented in a leaderboard (\authorc).

\subsection{Participants and learning context}
Approved by the Human Ethics Committee at the University of Queensland, this study was conducted within an introductory web design course at the same institution. The cohort consisted of 425 students, including 172 international and 253 domestic students from various disciplines, with 404 undergraduates and 21 graduate students. The gender distribution was 321 males and 104 females, with ages ranging from 17 to 56 years (average age: 20.9 years). Use of RiPPLE was embedded into the assessment design of the course, contributing 10\% of the final grade using four rounds of tasks and activities. Each round included the tasks of creating one resource, providing peer feedback on five resources and engaging in drill and practice study sessions that
at least include 10 of the resources created by their peers.

\subsection{Conditions}

\begin{figure}[htbp]
    \centering
    \includegraphics[width=.9\textwidth]{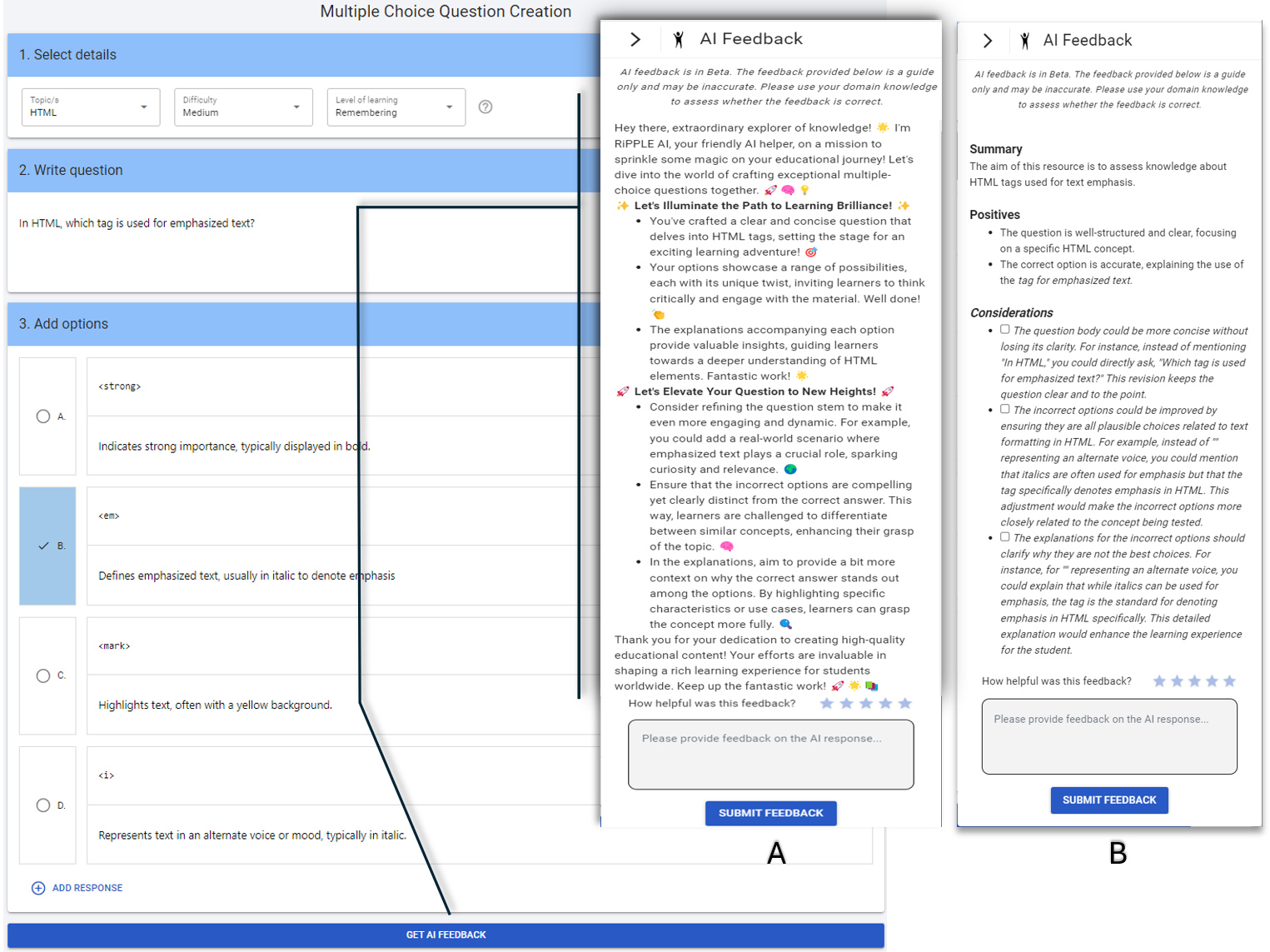}
    \caption{AI feedback examples for the Experimental Group (A) and Control Group (B) during resource creation}
    \label{fig:comparison}
\end{figure}

In this study, feedback prompts were carefully designed to examine the influence of emotionally enriched feedback on student responses. The control group received straightforward, constructive feedback with a neutral tone. In contrast, feedback for the experimental group was augmented with motivational phrases, praise, and emojis, aimed at creating a supportive and positive emotional environment (Figure \ref{fig:comparison}, Table\ref{tab:feedback_comparison}). These feedback elements were not merely decorative; emojis were strategically used as visual cues to enhance clarity and foster emotional connections \citep{Erle2021Emojis, Bai2019A}. Moreover, the feedback capitalized on social norms to set higher standards and promote peer accountability \citep{caraban201923}, thus nurturing a sense of community and positive social interactions that are essential for increased engagement and mitigating negative emotions \citep{pekrun2006control, frondozo2023feedback}.

\begin{table}
\centering
\caption{Comparison of Feedback Design Between Control and Experimental Groups}
\label{tab:feedback_comparison}

\renewcommand{\arraystretch}{1.5}
\footnotesize{
\begin{tabular}{>{\raggedright\arraybackslash}m{.15cm}| >{\raggedright\arraybackslash}m{2.3cm} >{\raggedright\arraybackslash}m{4.5cm} >{\raggedright\arraybackslash}m{6.5cm}}
\hline
& \textbf{Feedback Aspect/Section} & \textbf{Control Group} & \textbf{Experimental Group} \\
\hline
\multirow{5}{*}{\rotatebox{90}{\textbf{Aspects}}} & \textbf{Tone} & Neutral & Enthusiastic and supportive, using positive reinforcements and relevant emojis. \\
& \textbf{Content Focus} & Detailed and constructive criticism without emotional elements. & Similar content focus as control but includes motivational language and praise to enhance engagement. \\
& \textbf{Feedback Style} & Emphasis on actionable and specific feedback. Examples: "Avoid using loaded language in the options and explanations." & Maintains actionable and specific feedback but with an uplifting and fun tone, incorporating emojis. Examples: "Great job including explanations for each option! \includegraphics[height=10pt]{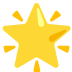}" \\
& \textbf{Use of Emojis} & None used. & Strategic use to enhance emotional connection and tone of the feedback. \\
& \textbf{Feedback  }(Word Count) & Mean = 190.9 & Mean = 278.6 \\

\hline
\multirow{5}{*}{\rotatebox{90}{\textbf{Sections}}} & \textbf{Introduction} & Not applicable & \textbf{Friendly Introduction:} Engages with a warm and inviting tone, setting a positive stage for feedback. \newline \textbf{Example:} "Hey there! \includegraphics[height=10pt]{emojis/Star.png} I'm RiPPLE AI, your friendly AI helper, ready to explore and enhance this question with you! \includegraphics[height=10pt]{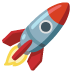} \includegraphics[height=10pt]{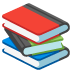} \includegraphics[height=10pt]{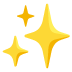}" \\
& \textbf{Summary} & \textbf{Summary as a section:} \newline \textbf{Example:} "The aim of this resource is to assess knowledge about HTML tags for creating hyperlinks." & \textbf{Summary Included in the Introduction:}  \newline \textbf{Example:} "Let's evaluate your resource: The question clearly tests HTML attributes related to frame loading." \\
& \textbf{Positive} & \textbf{Positive:} Identifies strengths in the resource. \newline \textbf{Example:} "The question is clearly written and addresses a fundamental concept in HTML." & \textbf{\includegraphics[height=10pt]{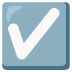} Positive:} Emphasizes strengths using motivational language and emojis. \newline \textbf{Example:} "Your question is clear and directly engages learners with a practical HTML scenario. This hands-on approach can enhance understanding and retention. Awesome job! \includegraphics[height=10pt]{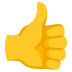}" \\
& \textbf{Constructive Feedback} & \textbf{Considerations:} \newline \textbf{Example:} "Consider rephrasing to 'Which HTML tag creates hyperlinks?' for clarity." & \textbf{\includegraphics[height=10pt]{emojis/rocket.png} Suggestions for Improvement:} \newline \textbf{Example:} "Try rephrasing the question to be more concise, such as 'Which value for the 'target' attribute loads a linked page into the topmost frame in HTML?'" \\
& \textbf{Last section} & Not applicable & \textbf{Acknowledgment of Contribution:} Expresses gratitude, encouraging ongoing improvement. \newline \textbf{Example:} "Thank you for your diligent work in crafting this educational content. \includegraphics[height=10pt]{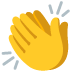}"  \newline\textbf{Closing Encouragement and Reminder:} Ends with motivational feedback and a peer review reminder. \newline \textbf{Example:} "Keep refining your materials; they will be reviewed by peers for further enhancement \includegraphics[height=10pt]{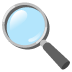}"  \\
\hline
\end{tabular}
}
\end{table}

\subsection{Experimental settings and data collection strategy}
We carried out an in-the-field randomised controlled experiment initially involving 425 participants. Students were randomly allocated to one of the two aforementioned conditions, experiencing either neutral feedback or emotionally enhanced feedback. In addition, during the platform sign-in stage, students were required to review a consent form to use their data for research and platform development purposes. Only the data of students who provided consent were obtained and considered for the data analysis. Overall, we used data from 381 students across the two groups authoring resources and performing peer feedback tasks, which are presented in Table \ref{tab:moderation_resources_updated}.

\begin{table}
\centering
\caption{Descriptive Statistics for Moderation Count and Total Resources Authored}
\label{tab:moderation_resources_updated}
\begin{threeparttable}
\begin{tabular}{lccccccc}
\toprule
\textbf{Metric} & \multicolumn{2}{c}{\textbf{N}} & \multicolumn{2}{c}{\textbf{Mean}} & \multicolumn{2}{c}{\textbf{Median}} & \textbf{Sum} \\
\cmidrule(r){2-3} \cmidrule(lr){4-5} \cmidrule(lr){6-7} \cmidrule(l){8-8}
 & \textbf{Ctrl} & \textbf{Exp} & \textbf{Ctrl} & \textbf{Exp} & \textbf{Ctrl} & \textbf{Exp} & \\
\midrule
Moderation Count & 192 & 189 & 12.63 & 12.70 & 15.00 & 15.00 & 5457 \\
Total Resources Authored & 192 & 189 & 4.09 & 4.04 & 4.00 & 4.00 & 1751 \\
\bottomrule
\end{tabular}
\end{threeparttable}
\end{table}

This investigation adhered to a rigorously structured methodology to ensure exhaustive data acquisition and analysis:
\begin{enumerate}
    \item \textbf{Participant Selection and Group Allocation:} Students enrolled in a web design course were randomly allocated to either the experimental or control group within the RiPPLE platform, ensuring a balanced distribution.
    \item \textbf{Survey Administration:}  Following the initial four-week period (Round 1), an online survey was distributed through Blackboard. Simultaneously, during the tutorial sessions, students were given the choice to participate in the survey, with a fifteen-minute window allocated for its completion.
    \item \textbf {Data Collection from RiPPLE:} At the conclusion of the course, data and log files were collected from the platform.
\end{enumerate}

\subsection{Data Analysis and Measures}

This study adopts a mixed-methods approach, utilizing both quantitative and qualitative analyses to assess the emotional and engagement responses of university students to AI feedback. 
We used SPSS version 29 for all statistical analyses and Qualtrics for survey data collection.

\textbf{RQ1. Engagement with Feedback}. In addressing RQ1, this study explores how AI-generated feedback influences student engagement on RiPPLE. Specifically, it measures the time between the receipt of feedback and the subsequent submission or moderation of resources as a proxy of time taken for the student to reflect on and action the suggested considerations. Outliers, defined as instances where students received feedback but did not engage further with the platform, were removed from the dataset—13 data points for creation and 4 for moderation—to ensure more accurate results  Due to the non-normal distribution of engagement times, we employed the non-parametric Mann-Whitney U test to compare engagement efficiency between the control and experimental groups.Additionally, to assess overall engagement, the study utilized three survey items adapted from \citep{kay2011evaluating}. These items specifically measure the appreciation of thematic feedback, the levels of student engagement, and the enjoyment derived from the feedback, thereby providing a comprehensive understanding of student interaction with the feedback process.

\textbf{RQ2. Usefulness perception}. 
for addressing RQ2, data were collected from RiPPLE where students engaged in creation or moderation tasks. After receiving AI-generated feedback, students had the opportunity to rate the quality of this feedback using a 1-5 star scale, as depicted in Figure \ref{fig:UML}. This rating system allowed us to directly assess their perceptions of the feedback's effectiveness. Given the categorical nature of the scale and the non-normal distribution of our data, the Mann-Whitney U test was employed.Additionally, 79 comments regarding the usefulness of AI feedback were collected from the RiPPLE platform. . 

\textbf{RQ3. Emotional response to feedback}. In response to RQ3, we conducted a detailed 41-item survey, measuring eight distinct emotions, each through four dimensions—Affective, Motivational, Physical, and Cognitive—using an adapted version of the AEQ-S \cite{bieleke2021aeq}. We also included five open-ended questions to gather qualitative insights, enriching our understanding of the quantitative data. Given the non-normal distribution of the data, the Mann-Whitney U test was utilized (Table \ref{tab:emotions-p.value}).
Out of the 103 students who participated in the survey, 49 from the control group and 54 from the experimental group, with 2 participants removed due to their exceptionally short completion times of 93 seconds or less, compared to the average survey completion time of 10 minutes, 70 responded to the optional qualitative questions.

 \textbf{RQ4. Impact on outcome.} For RQ4 data were collected from RiPPLE regarding the quality of resources created by both experimental and control groups. Each resource was evaluated and assigned a quality score ranging from 0 to 5. This multi-dimensional scoring approach, as detailed by \cite{darvishi2021employing}, 
 integrates student ratings, instructor evaluations, and sophisticated algorithmic assessments. Students provide initial ratings based on relevance and overall quality, while instructors contribute crucial moderation, ensuring the robustness of the final scores. Advanced algorithms, including Expectation-Maximization and graph-based trust propagation models, are used to refine these assessments and adjust for biases related to user reliability. A Shapiro-Wilk test confirmed the non-normality of the quality scores (p < 0.05), prompting the use of the Mann-Whitney U test to examine discrepancies in quality between the groups (Table \ref{tab:Resource_Quality}).

\textbf{Qualitative Analyses.}
The qualitative data from both RQ2 and RQ3 were synthesized and analyzed using Reflexive Thematic Analysis, offering a richer understanding of students' perceptions.\citep{Braun2022}. The transcripts were systematically analyzed by reading the materials twice and manually identifying the significant features that were coded. To enhance accuracy and comprehensiveness, the transcripts were reviewed by a second researcher, who reviewed the transcripts and made a second list of significant features. Together, the researchers compared their coded lists and clustered similar codes into initial themes and refined them through collaborative review. Each theme was distinctly named and defined to encapsulate its essence. Finally, the authors jointly wrote the analysis, presenting evidence and interpretations.

\section{Results} \label{sec:Results}

\subsection{Impact on Student Feedback Engagement (RQ1)}

Table~\ref{tab:time_difference_groups} presents the engagement metrics for AI feedback during the creation and moderation tasks. These metrics include the duration (in seconds) that students interacted with the feedback. The \(N\) values represent the number of resources or moderation instances evaluated. In the creation task, the control group exhibited a median engagement time of (\(Mdn = 76.545\), \(N = 707\)), while the experimental group had a median time of (\(Mdn = 70.6865\), \(N = 696\)). This difference was statistically significant (\(p = 0.026\)) with a small effect size (\(|r| = 0.059\)). For the moderation task, the engagement times were nearly identical, with the control group at (\(Mdn = 3.450\), \(N = 2709\)) and the experimental group at (\(Mdn = 3.464\), \(N = 2671\)), showing no statistically significant difference (\(p = 0.920\)). Additionally, further results on engagement from the survey data revealed no significant differences between the control group ((\(Mdn = 3.29\), \(N = 50\)) and the experimental group ((\(Mdn = 3.25\), \(N = 51\)), (\(p = 0.823\))).

\begin{table}
    \centering
    \caption{AI Feedback Engagement During Creation and Moderation Tasks}
    \label{tab:time_difference_groups}
    \begin{threeparttable}
        \begin{tabular}{lcccccc}
            \toprule
            \textbf{Task} & \multicolumn{2}{c}{\textbf{N}} & \multicolumn{2}{c}{\textbf{Median (s)}} & \textbf{\(p\)-value} & \textbf{Effect Size \( |r| \)}\\
            \cmidrule(r){2-3} \cmidrule(lr){4-5}
            & \textbf{Ctrl} & \textbf{Exp} & \textbf{Ctrl} & \textbf{Exp} & & \\
            \midrule
            Creation & 707 & 696 & 76.545 & 70.6865 & 0.026 & 0.059 \\
            Moderation & 2709 & 2670 & 3.450 & 3.464 & 0.920 & 0.00138 \\
            \bottomrule
        \end{tabular}
        \begin{tablenotes}
            \item \textit{Note: Median times converted from ms to seconds.}
        \end{tablenotes}
    \end{threeparttable}
\end{table}

\subsection{Perceptions of the Benefits of AI-Generated Feedback (RQ2)}

The analysis of student perceptions regarding the effectiveness of AI-generated feedback, detailed in Table~\ref{tab:ai_feedback}, shows noticeable differences between the control and experimental groups. In the moderation task, the control group evaluated the feedback with a \(M_{rank} = 139.79\), (N=91), while the experimental group gave a higher \(M_{rank} = 190.74\), (N=103). This difference was statistically significant (\(p<0.001\)) and had a moderate effect size (\(r=0.289\)).

\begin{table}
    \centering
    \caption{Students’ Ratings on AI-Generated Feedback for Moderation and Authored Resources Tasks}
    \label{tab:ai_feedback}
    \begin{threeparttable}
        \begin{tabular}{lcccccccccc}
            \toprule
            \textbf{Task} & \multicolumn{2}{c}{\textbf{N}} & \multicolumn{2}{c}{\textbf{Mean}} & \multicolumn{2}{c}{\textbf{Median}} & \multicolumn{2}{c}{\textbf{Std. Dev.}} & \textbf{\(p\)-value} & \textbf{Effect Size \( |r| \)} \\
            \cmidrule(r){2-3} \cmidrule(lr){4-5} \cmidrule(lr){6-7} \cmidrule(lr){8-9}
            & \textbf{Ctrl} & \textbf{Exp} & \textbf{Ctrl} & \textbf{Exp} & \textbf{Ctrl} & \textbf{Exp} & \textbf{Ctrl} & \textbf{Exp} & & \\
            \midrule
            Moderation & 91 & 103 & 3.97 & 4.41 & 4.00 & 4.00 & 1.00 & 1.02 & 0.001** & 0.289 \\
            Creation & 56 & 44 & 4.16 & 4.57 & 4.50 & 5.00 & 1.047 & 0.728 & 0.040* & 0.206 \\
            \bottomrule
        \end{tabular}

    \end{threeparttable}
\end{table}

Similarly, in the creation task, there were significant differences in how feedback was rated by the two groups. The control group's \(M_{rank} = 43.89\) (N=54), compared to the experimental group's higher \(M_{rank} = 54.43\) (N=42). This result was also statistically significant (\(p=0.040\))  and demonstrated a small to medium effect size (\(r=0.206\)). 

\subsection{Effects of AI Feedback on Emotional Responses (RQ3)}

The survey findings, detailed in Table~\ref{tab:emotions-p.value}, provide insights into the emotional impacts of AI feedback on students. In terms of enjoyment, both the control and experimental groups reported moderate levels, with median responses of 3.31 and 3.12 respectively, indicating no statistically significant difference (\(p = 0.984\), effect size \(|r| = 0.002\)). Similarly, levels of hope were comparable between the groups, with the control group reporting a median of 3.38 and the experimental group 3.29, suggesting no significant difference (\(p = 0.773\), effect size \(|r| = 0.029\)). The median levels of pride were also closely matched across groups, with 3.42 in the control and 3.31 in the experimental group, showing no statistical significance (\(p = 0.854\), effect size \(|r| = 0.018\)).

\begin{table}
  \centering
  \caption{Statistical Analysis of Emotional Responses: Control vs. Experimental Groups}
  \label{tab:emotions-p.value}
  \begin{threeparttable}
    \begin{tabular}{lcccc}
    \toprule
    \textbf{Variable} & \multicolumn{2}{c}{\textbf{Median}} & \textbf{\( p \)-value} & \textbf{Effect Size \( |r| \)}\\
    \cmidrule(r){2-3}
          & \textbf{Ctrl}  & \textbf{Exp} & & \\
    \midrule
    Enjoyment & 3.31  & 3.12  & .984 & 0.002 \\
    Hope  & 3.38  & 3.29  & .773 & 0.029 \\
    Pride & 3.42  & 3.31  & .854 & 0.018 \\
    Anger & 2.61  & 1.89  & .013* & 0.246 \\
    Anxiety & 1.82  & 1.62  & .235 & 0.118 \\
    Shame & 1.71  & 1.46  & .179 & 0.134 \\
    Hopelessness & 1.73  & 1.51  & .452 &0.075 \\
    Boredom & 2.81  & 2.47  & .212 & 0.124 \\
    \bottomrule
    \end{tabular}

  \end{threeparttable}
\end{table}

Conversely, the experimental group exhibited significantly lower levels of anger, with a median of 1.89 compared to 2.61 in the control group, demonstrating a notable reduction (\(p = 0.013^*\), effect size \(|r| = 0.246\)). Anxiety levels were low in both groups (medians of 1.82 and 1.62), with no statistically significant differences observed (\(p = 0.235\), effect size \(|r| = 0.118\)). Moreover, other negative emotions such as shame and hopelessness reported similarly low levels, with the control group medians at 1.71 and 1.73, and the experimental group at 1.46 and 1.51, respectively (\(p > 0.05\), effect size \(|r| = 0.134\) for shame and \(|r| = 0.075\) for hopelessness). Additionally, boredom levels were also measured, with the control group reporting a median of 2.81 compared to 2.47 in the experimental group. Although this difference was not statistically significant (\(p = 0.212\)), the effect size (\(|r| = 0.124\)) suggests a small but notable lower level of boredom in the experimental group.\\

\begin{table}
    \centering
    \caption{Analysis of Four Dimensions Measuring Anger}
    \label{tab:anger_motivation}
    \begin{threeparttable}
        \begin{tabular}{p{4cm}p{2.5cm}cccccc}
            \toprule
            \textbf{Question} & \textbf{Dimension} & \multicolumn{2}{c}{\textbf{Median}} & \multicolumn{2}{c}{\textbf{Std. Dev.}} & \textbf{\(p\)-value} & \textbf{Effect Size \( |r| \)} \\
            \cmidrule(r){3-4} \cmidrule(l){5-6}
            & & \textbf{Ctrl} & \textbf{Exp} & \textbf{Ctrl} & \textbf{Exp} & & \\
            \midrule
            I feel frustrated when the AI feedback points out errors or shortcomings in my task. & Affective & 2.956 & 2.526 & 1.164 & 1.041 & 0.062 & 0.186 \\
            \midrule
            I get annoyed about having to address AI suggestions for improvements on my task. & Motivational & 3.223 & 2.582 & 1.141 & 1.082 & 0.005* & 0.279 \\
            \midrule
            The AI feedback angers me to the point where I feel like throwing the keyboard out of the window. & Physical & 2.142 & 1.890 & 1.248 & 1.209 & 0.153 & 0.141 \\
            \midrule
            Receiving critical AI feedback can trigger restlessness or discomfort. & Cognitive & 2.313 & 2.048 & 1.136 & 1.120 & 0.132 & 0.149 \\
            \midrule
        \end{tabular}

    \end{threeparttable} 
\end{table}
Furthermore, the analysis across four dimensions measuring distinct emotional responses uncovered a particularly notable result within the dimension of anger related to motivation, as illustrated in Table~\ref{tab:anger_motivation}. Specifically, the findings indicate that students in the control group exhibited significantly higher levels of annoyance—an indicator of demotivation—when prompted to address AI-generated suggestions for improvements on their tasks. This difference was not only statistically significant (\(p < 0.005\)), but also exhibited a moderate effect size (\(|r| = 0.279\)).

\subsection{Effect on Quality of Student Work (RQ4)}

In response to (RQ4), the analysis revealed that there was no statistically significant difference in the quality scores between the experimental and control groups (\(p\)= 0.470), with both groups achieving a score of \(Mdn = 4.10\) as shown in Table~\ref{tab:Resource_Quality}.

\begin{table}
    \centering
    \caption{Assessment of Resource Creation Task Quality}
    \label{tab:Resource_Quality}
    \begin{threeparttable}
        \begin{tabular}{lccccccccccc}
            \toprule
            \textbf{Mertic} & \multicolumn{2}{c}{\textbf{N}} & \multicolumn{2}{c}{\textbf{Mean}} & \multicolumn{2}{c}{\textbf{Median}} & \multicolumn{2}{c}{\textbf{Std. Dev.}} & \multicolumn{2}{c}{\textbf{Shapiro-Wilk Sig.}} & \textbf{ \(p\)-value} \\
            \cmidrule(r){2-3} \cmidrule(lr){4-5} \cmidrule(lr){6-7} \cmidrule(lr){8-9} \cmidrule(lr){10-11}
            & \textbf{Ctrl} & \textbf{Exp} & \textbf{Ctrl} & \textbf{Exp} & \textbf{Ctrl} & \textbf{Exp} & \textbf{Ctrl} & \textbf{Exp} & \textbf{Ctrl} & \textbf{Exp} & \\
            \midrule
            Resource Quality & 744 & 751 & 4.107 & 4.076 & 4.100 & 4.100 & 0.5693 & 0.5966 & <.001 & <.001 & 0.470 \\
            \bottomrule
        \end{tabular}

    \end{threeparttable}
\end{table}

\subsection{Qualitative Results}

\textbf{Theme 1: Positive Impact on Learning Experience.} Nearly half of the responses (49\%) indicated that the AI feedback was perceived positively, noting that it enhanced their learning experience by providing valuable insights and increasing engagement. For instance, one student said, \textit{"Made me more engaged in the tasks"} (Experimental group (Exp)), while another commented, \textit{"It gave me precise advice on which part I am doing wrong. According to its feedback, I could finish my task more quickly and correctly"} (Control group (Ctrl)). Students highlighted how the feedback made them more motivated to improve, with comments such as \textit{"It makes me more motivated to improve my questions and more engaging to do RiPPLE"} (Exp) and \textit{"It has allowed me to improve my questions before release, which increases my confidence"} (Exp). There was widespread agreement with the feedback, as many students felt that the AI's suggestions were relevant and useful for task improvement.

\textbf{Theme 2: Neutral or Minimal Impact.} Around 9\% of responses indicated that the AI feedback had little to no impact on the students' learning experience. Comments such as \textit{"Not affected my engagement"} (Ctrl) and \textit{"I generally don't feel anything towards the AI feedback"} (Ctrl) reflect this sentiment. These responses suggest that the feedback neither enhanced nor hindered their progress, offering minimal support in their academic tasks.

\textbf{Theme 3: Emotional Impact of AI Feedback.} 21\% of the comments reflected students' emotional reactions to the AI feedback, which ranged from positive to negative.

Positive emotions like satisfaction and confidence were reported by 11\% of the students' responses. For example, one student expressed, \textit{"It made me feel sort of relieved as it pointed out things that I could improve on to make good feedback"} (Ctrl). Another noted, \textit{"I feel a bit confident if it gives me a good comment"} (Ctrl). Some students appreciated the non-judgmental nature of the AI, with one stating, \textit{"I like how the feedback is not from a person and lacks the judgement associated with that. I find the feedback more easily absorbed as I don't feel critically judged"} (Exp). Another student mentioned, \textit{"The AI feedback made me feel positive about my feedback when it gave me positive feedback as well"} (Exp).

Negative emotions were reported by 10\%, particularly in response to delays in feedback or perceived inaccuracies. One student mentioned, \textit{"I find much less enjoyment doing the RiPPLE tasks as I get annoyed when waiting for the AI feedback to be typed out"} (Exp), reflecting frustration with the system's response time. Another student expressed dissatisfaction, stating, \textit{"It hasn't always been accurate and hasn't always been able to pick up on the tone that I was going for with my feedback/my questions"} (Ctrl).

\textbf{Theme 4: Issues with Feedback Quality and Relevance.} Nearly 18\% of responses raised concerns about the quality and relevance of the feedback. Redundancy was a common complaint, with students stating that the AI often repeated suggestions or failed to provide new insights. One student remarked, \textit{"The AI is asking me to do things I have already done"} (Ctrl). Lengthy feedback was also a source of frustration, with comments like \textit{"The feedback is a bit too long"} (Exp) and \textit{"It was a lot to read so I did not read all of it"} (Ctrl). Additionally, misalignment between the feedback and students' input caused confusion. For instance, a student stated, \textit{"It has not always been accurate and hasn't always been able to pick up on the tone that I was going for with my feedback/my questions"} (Ctrl). Another student mentioned, \textit{"Sometimes it suggests excessive things that would not be required for the level of the task"} (Ctrl). Instances where the AI overlooked key user input were also noted, contributing to confusion and mistrust. One student reported, \textit{"It never seems to add anything of value, sometimes even missing points that you have already made"} (Ctrl).

\textbf{Theme 5: Technical Challenges.}

Only 3\% of responses mentioned technical difficulties with the AI feedback system. These operational issues included delays in feedback generation and system failures. One student noted, \textit{"I had to wait for the feedback generation, and it is usually a long paragraph"} (Exp), while another reported, \textit{"Sometimes it fails to work"} (Ctrl). These challenges contributed to a sense of frustration and diminished the overall effectiveness of the feedback.

\section{Discussion} \label{sec:Dis}
Learnersourcing promotes socio-cultural and socio-cognitive learning activities that support students' cognitive growth (see Section~\ref{sec:system}). The socio-cognitive theoretical perspective positions students as agents who interact with their environments to construct and develop their knowledge. Hence, in learning platforms employing learnersourcing, for students to exercise agency, they are required to develop SRL skills \citep{schunk_social_1989}. From the socio-cultural viewpoint, SRL skills can be supported through scaffolding strategies such as monitoring approaches \citep[see][]{azevedo_lessons_2022}. The RiPPLE platform employed in this study has previously investigated the efficacy of various learning scaffolds to support SRL processes. For instance, \authord et al. proposed and evaluated four SRL scaffolding strategies against a control group: (1) self-monitoring, (2) planning, (3) self-assessing, and (4) a strategy combining all three of these approaches. One of the study's important findings was that students perceived the four strategies differently, with more negative emotions associated with the condition that involved more complex learning experiences (the fourth condition). This finding highlighted the importance of maintaining student motivation and emotion while engaging with the platform. Therefore, building on our previous endeavours and the recent feature added to the platform that utilizes LLMs to provide students with automated feedback on their learning products, we addressed the emotional aspect of learning using the CVT of achievement emotions. Our study not only contributes to the development and enhancement of RiPPLE but also partially responds to the call made by \cite{tabak_research_2018}, questioning whether technological adaptive scaffolding is inherently limited by the absence of human affective components. In what follows, we discuss our findings in relation to the RQs presented in Section~\ref{sec:intro}.

\subsection{RQ1: Engagement with feedback}
The literature on peer assessment identifies several forms of student engagement with feedback provided by peers, including cognitive, behavioural, and emotional engagement \citep{jin_effects_2022}. While previous studies have shown that students are willing to engage with and benefit from peer feedback, they may also encounter some challenges \citep{cheng_exploring_2014}. RQ1 aimed to examine student engagement with emotionally enriched LLMs-based written feedback from a behavioural perspective. We measured the time students spent reading and reflecting on the LLMs-based feedback they received while authoring or reviewing learning content. Regarding the peer review task, the finding revealed no significant difference between the groups. Regarding the content authoring task, the findings revealed a notable difference in the time spent between the control and experimental groups, with the control group demonstrating longer engagement time with the feedback. Although there was a statistically significant difference in engagement time between the two groups, the effect size was small. This small effect size is also reflected in the minor difference in average time (i.e., 6 seconds). At first glance, this finding might appear unexpected because the experimental group received longer comments, which are assumed to require longer reading time.

However, this finding can be explained in the context of previous research and from the perspective of the CVT. First, \cite{Cloude_negative_2021} found that fluctuations in negative deactivating emotions over six-time points were negatively associated with the time spent utilizing cognitive strategies. Hence, the students in the experimental group might have had unstable emotions due to the added emotional features in the feedback, where different parts of the feedback conveyed different emotions through emojis and praises. Second, CVT implies that emotions are associated with motivation and cognition \citep{pekrun2006control}. Students experiencing negative emotions might not focus on the task at hand and consume cognitive resources \citep{pekrun2006control}. In our study, this might have been the case with the control group, where they might have had negative emotions in response to the feedback comments. Third, central to CVT are the control and value appraisals based on which students perceive their control over success and failures and the value of outcomes. Hence, in the case of the experimental group, the students might have felt satisfied when evaluating the feedback, believing that their initial efforts were adequate due to the emotional features that praised their effort \citep{kakinuma2022praise}.

\subsection{RQ2: Perceptions of feedback usefulness}
While there are currently several discussions on the effectiveness of incorporating LLMs in the learning process to support students \citep{kasneci_chatgpt_2023}, how students perceive LLMs-based feedback is still under investigation. Examining students’ perceptions is important because it can impact various learning factors such as online learning satisfaction, intrinsic motivation, and feedback processing and revision \citep{noroozi_does_2024}. Recent research findings are mixed: some studies found that students perceived human feedback (e.g., peers) as more useful than LLMs-based feedback \citep[e.g.,][]{olga2023generative}, while other research found that students’ preferences were split between both sources of feedback \citep[e.g.,][]{escalante_ai-generated_2023}. However, complementing human feedback with LLMs was perceived as more useful than human or LLMs-based feedback alone. RQ2 contributes to this knowledge by examining student perceptions of the usefulness of emotionally enriched LLMs-based written feedback. We measured students’ perceptions using helpfulness ratings presented after they received LLMs-based written feedback on content authoring and peer review tasks. The findings revealed a significant difference between the two groups, with students who received emotionally enriched feedback finding the LLMs-based feedback more useful than their peers in the control group.

The RQ2 findings align with the interpretation of the RQ1 findings, where we posited that the students in the experimental group were more satisfied with the outcomes of their work on content authoring. However, this finding might raise concerns regarding feedback enactment; one might assume that students who received feedback complemented with praise and emojis are less motivated to revise their initial work \citep[see][]{brophy_teacher_1981}. We argue that students who receive emotionally enriched feedback might be more willing to reengage with learning tasks. For instance, a student might complete a content authoring task and follow it with another content authoring task on the learning platform; hence, spending less time on the task at hand but more time on the platform than students who do not receive emotionally enriched feedback. This argument is supported by research in Human-Computer Interaction. For example, \cite{obrien_what_2008} described user engagement as a cyclic process in which positive prior experiences with an application increase the likelihood of reengaging with the application.

\subsection{RQ3: Emotional response to feedback}
The CVT of achievement emotions can provide valuable insights into emotion generation, emotion sources, and associated processes in complex, intelligent learning environments \citep[e.g.,][]{azevedo_lessons_2022}. One of the essential assumptions of CVT is that scaffolding the regulation of value and control appraisals can indirectly support student emotional development \citep{pekrun2006control}. Hence, RQ3 examined student emotions by utilizing the CVT of achievement emotions. Specifically, we followed the CVT assumptions to scaffold positive emotions in both content authoring and peer review tasks, then compared the scaffolded group (experimental) against the control group across several emotional dimensions: enjoyment, hope, pride, anger, anxiety, shame, hopelessness, and boredom. We measured these emotions using a survey instrument after an extended period of time, unlike in RQ2. The findings demonstrated relatively high levels of positive emotions with no significant differences between the groups. Conversely, the negative emotion dimensions revealed a different story, where students experienced low levels of negative emotions with the exception of anger and boredom, which were relatively moderate. More importantly, students in the control group demonstrated a higher level of anger than the experimental group, particularly in response to the statement, ``I get annoyed about having to address AI suggestions for improvements on my task.’’ This finding indicates that our emotional scaffolding technique could lower student anger.

However, this does not necessarily mean our intervention increased or maintained students’ motivation in the experimental group when receiving the LLMs-based feedback. As we described earlier, the students in the experimental group might have had low motivation to revise their initial work due to their feeling of satisfaction. Hence, linking the findings of RQ3 with those of RQ1 and RQ2, we can assume that, in alignment with previous research findings, the response behaviors of our students were associated with their emotions \citep[e.g.,][]{cheng_exploring_2014}. The control group experienced feelings of anger when enacting the LLMs-based feedback, but they attempted to address it while experiencing this negative emotion, which led them to consume more cognitive resources and spend more time on the task at hand. Conversely, the experimental group felt more satisfaction and less anger, so they spent less time. This result can also be explained from an attribution perspective, which states that following task completion, students might experience emotional reactions regarding task outcomes (e.g., sadness or happiness in failure and success situations) and reflect on these outcomes. These emotional reactions might further generate complex emotions such as anger and pride \citep{boekaerts_chapter_2000}.

\subsection{RQ4: Quality of student work}
Students’ learnersourced product quality can indirectly reflect various learning phenomena. Generally, high-quality products (i.e., authored learning resources and written reviews) can indicate cognitive engagement with the tasks, thus, learning occurs. For instance, in the case of content authoring, students could engage in high-level cognitive processes reflected in Bloom’s taxonomy \citep[e.g.,][]{moore_assessing_2022}. In the context of LLMs-based feedback, such engagement could occur because students reflect on the feedback they receive. Hence, high-quality products might indicate that students have developed feedback literacy, where they value their role in the feedback process. For instance, students could feel responsible for taking action to incorporate the received feedback to improve the quality of their work \citep[see][]{carless_development_2018}. RQ4 examined the impact of incorporating emotions into LLMs-based feedback. We used the quality rating provided by RiPPLE to determine each learning product's quality. The findings demonstrated that both the control and experimental groups generated learning products of equivalent quality, which might indicate that feedback uptake was similar in both groups. This finding is surprising, as students in the control group had more behavioural engagement (see the discussion of RQ1). A study on the effect of student engagement on writing tasks by \cite{jin_effects_2022} found that behavioural engagement was associated with writing performance. However, in our study, the higher level of behavioural engagement in the control group was accompanied by negative emotions (i.e., anger), which might explain why their performance was not higher than that of the experimental group.

The findings of RQ4 contribute to the line of research concerning the effect of LLMs-based feedback on learning. Recent studies have shown how LLMs-based feedback can complement student-generated content to improve the quality of their work \citep[][]{moore_assessing_2022}. Additionally, other studies have demonstrated that LLMs-based feedback is equivalent to human feedback in terms of learning outcomes \citep[e.g.,][]{escalante_ai-generated_2023}. Our study’s findings imply that the effect of emotionally enriched LLMs-based feedback on learning outcomes is complex. This aligns with previous research on the effect of positive and negative emotions on effort and learning \citep[see][]{schunk_self-regulated_2005}. However, linking the findings of RQ4 to those of RQ2 and RQ3, what is significant about the findings of RQ4 is that our intervention could maintain students' emotional well-being without compromising their learning outcomes.

\subsection{Implications for Educational Practice}
This study's findings underscore the importance of integrating emotional elements into AI-driven feedback systems for enhancing the educational experience. Emotionally enriched AI feedback can substantially improve students' emotional well-being and their perception of feedback, which are pivotal for fostering a conducive learning environment. Our results advocate for a balanced approach that considers both the cognitive and emotional aspects of student learning, suggesting that educational technologies should not merely focus on academic performance but also on enhancing the overall educational journey. Implementing such feedback mechanisms could lead to more engaged and motivated learners, potentially leading to improved academic outcomes over time. Drawing from the insights gained in our study, we outline key design guidelines that can help in crafting more effective and emotionally resonant AI feedback systems:

\begin{itemize}
    \item \textbf{Tailor Emotional Tone to Task Type:} Our findings indicate that the type of task (creation vs. moderation) influences how students engage with feedback. Design feedback that adjusts its tone and emotional content based on the task's nature to enhance engagement and effectiveness. For instance, creation tasks may benefit from more motivational and uplifting feedback to encourage creativity and exploration.

    \item \textbf{Optimize Feedback Length and Clarity:} Feedback effectiveness was influenced by its perceived clarity and brevity. Ensure that AI-generated feedback is concise yet comprehensive, avoiding overly verbose responses that may lead to disengagement or overwhelm the learner.

    \item \textbf{Incorporate Visual Emotional Cues Appropriately:} Given the positive reception to the use of emojis in feedback, they should be strategically employed to highlight key points and enhance understanding. However, their use must be contextually appropriate and should not detract from the feedback's main message.

    \item \textbf{Address Negative Emotions Directly:} Feedback should be designed to recognize and mitigate negative emotions like anger, anxiety or boredom, which can hinder learning.

    \item \textbf{Use Positive Reinforcement to Mitigate Negative Feedback:} When delivering negative feedback or pointing out errors, it's beneficial to couple this with positive reinforcement to maintain motivation and soften the impact of criticism .

\end{itemize}

These guidelines are based directly on the quantitative and qualitative findings of our study, emphasizing a nuanced approach to integrating emotional intelligence into AI-driven educational feedback systems. By adhering to these principles, educators and technologists can better support students' emotional and academic growth in digital learning environments.

\subsection{Future work}
This study provides valuable insights into the effectiveness of emotionally enriched AI feedback within educational settings, yet it encountered several challenges.The absence of tools for real-time emotional measurement within the RiPPLE platform limits the capability to offer personalized feedback based on the immediate emotional states of students, thereby restricting the feedback's responsiveness and specificity. Technical constraints inhibited the precise measurement of how students engaged with and applied the feedback provided, restricting our understanding of the practical usage and direct impact of the feedback on students' learning processes.
Building upon the findings of this study, future research should aim to enhance the design and functionality of AI-driven emotional feedback systems. A key aspect of this future work could involve measuring emotions immediately after using RiPPLE and again three weeks later to assess both immediate and sustained effects. Additionally, conducting a three-month survey on emotions would provide valuable insights into the long-term impacts of emotionally enriched feedback. Developing advanced tools to track real-time engagement and application of feedback will provide deeper insights into its effectiveness, helping to understand how students use feedback in their learning process and how it impacts their performance. Further investigations should explore how different formats and styles of emotionally enriched feedback impact student engagement and feedback uptake. This exploration should aim to refine feedback mechanisms to maximize engagement, making the feedback not only more emotionally resonant but also more compelling and actionable for students. Additionally, implementing sophisticated analytical tools that can process real-time data to provide feedback that is not only emotionally intelligent but also contextually adapted to the learning environment and individual learner profiles will significantly improve the effectiveness of emotionally enriched feedback systems, making them more adaptable, responsive, and ultimately more supportive of students' educational and emotional needs.

\section{Conclusion}
This research highlights the significant potential of integrating emotionally enriched feedback via generative AI in educational settings, particularly focusing on enhancing the emotional aspects of student feedback encounters. The study employed a randomized controlled trial involving a significant cohort of first-year engineering students, assessing the impact of AI-generated feedback embellished with motivational elements on student engagement and emotional responses.

The experimental group, which received feedback augmented with positive reinforcements like praise and emojis, reported a noticeable improvement in their emotional well-being and perceived the feedback as more beneficial compared to the neutral feedback received by the control group. However, the study also revealed that while the emotional enhancements did not significantly impact the overall level of engagement with feedback or improve the quality of student work, they did foster a more positive feedback environment.

These findings suggest that while the integration of emotional elements into AI-driven feedback does not directly correlate with enhanced academic performance, it plays a crucial role in improving the quality of the educational experience by reducing negative emotions and enhancing student receptivity to feedback. This insight is particularly relevant in digital learning environments where lack of physical cues can often lead to emotional disconnect and decreased student motivation.

Looking forward, this study highlights the need for further exploration into the long-term effects of emotionally enriched AI feedback on academic outcomes. It also suggests a potential expansion of such feedback systems across various educational settings to comprehensively understand its effectiveness. Moreover, future research could focus on refining AI algorithms to better address the nuanced emotional and cognitive needs of students, potentially enhancing both engagement and academic success.

This research not only contributes to the evolving landscape of educational technology but also highlights the importance of addressing emotional factors in designing effective educational feedback systems. By fostering a more empathetic and supportive feedback environment, educational technologies can better support student learning and emotional health, thereby enhancing the overall educational journey.

\bibliographystyle{unsrtnat}
\bibliography{references}  %%% Uncomment this line and comment out the ``thebibliography'' section below to use the external .bib file (using bibtex) .

%%% Uncomment this section and comment out the \bibliography{references} line above to use inline references.
% \begin{thebibliography}{1}

% 	\bibitem{kour2014real}
% 	George Kour and Raid Saabne.
% 	\newblock Real-time segmentation of on-line handwritten arabic script.
% 	\newblock In {\em Frontiers in Handwriting Recognition (ICFHR), 2014 14th
% 			International Conference on}, pages 417--422. IEEE, 2014.

% 	\bibitem{kour2014fast}
% 	George Kour and Raid Saabne.
% 	\newblock Fast classification of handwritten on-line arabic characters.
% 	\newblock In {\em Soft Computing and Pattern Recognition (SoCPaR), 2014 6th
% 			International Conference of}, pages 312--318. IEEE, 2014.

% 	\bibitem{hadash2018estimate}
% 	Guy Hadash, Einat Kermany, Boaz Carmeli, Ofer Lavi, George Kour, and Alon
% 	Jacovi.
% 	\newblock Estimate and replace: A novel approach to integrating deep neural
% 	networks with existing applications.
% 	\newblock {\em arXiv preprint arXiv:1804.09028}, 2018.

% \end{thebibliography}

\end{document}